\documentclass[preprint2]{aastex}

\usepackage{graphicx}

\shorttitle{Search for Interstellar Adenine}
\shortauthors{Chakrabarti et al.}

\begin{document}


\title{Search for Interstellar Adenine}


\author{Sandip K. Chakrabarti}
\affil{ S. N. Bose National Centre for Basic Sciences, Salt Lake,
              Kolkata- 700098, India}
\affil{                    \&}
\affil{Indian Centre for Space Physics, Chalantika 43, Garia Station Road,
             Kolkata- 700084, India}

\author{Liton Majumdar}
\affil{Indian Centre for Space Physics, Chalantika 43, Garia Station Road,
             Kolkata- 700084, India}
\author{Ankan Das}
\affil{Indian Centre for Space Physics, Chalantika 43, Garia Station Road,
             Kolkata- 700084, India}

\author{Sonali Chakrabarti}
\affil{Maharaja Manindra Chandra College, 20 Ramakanto Bose Street,
             Kolkata- 700003, India}
\affil{                    \&}
\affil{Indian Centre for Space Physics, Chalantika 43, Garia Station Road,
             Kolkata- 700084, India}
%



\begin{abstract}
It is long debated if pre-biotic molecules are indeed present in the interstellar medium.
Despite substantial works pointing to their existence, pre-biotic molecules are yet to be 
discovered with a complete confidence. In this paper, our main aim is to
study the chemical evolution of interstellar adenine under various circumstances.
We prepare a large gas-grain chemical network by considering various pathways for the 
formation of adenine. \citet{maju12} proposed that in the absence of 
adenine detection, one could try to trace two precursors of adenine, namely,
$\mathrm{HCCN}$ and $\mathrm{NH_2CN}$. Recently \citet{merz14},
proposed another route for the formation of adenine in interstellar condition.
They proposed two more precursor molecules. But it was not verified by any 
accurate gas-grain chemical model. Neither was it known if the production rate would be high or low.
Our paper fills this important gap. We include this new pathways
to find that the contribution through this pathways for the formation of Adenine 
is the most dominant one in the context of interstellar medium. We propose that observers may look for the two 
precursors ($\mathrm{C_3NH}$ and $\mathrm{HNCNH}$) in the interstellar media which are equally important 
for predicting abundances of adenine. 
We perform quantum chemical calculations to find out spectral properties of adenine and its two new 
precursor molecules in infrared, ultraviolet and sub-millimeter region. Our present study would 
be useful for predicting abundance of adenine.
\end{abstract}
\maketitle
\section{Introduction}           
According to the Cologne Database for Molecular Spectroscopy \citep{mull01,mull05}, around
$180$ molecules have been detected in the interstellar medium (ISM) 
or in circumstellar shells. Among these, severals species are organic in nature.
Existence of these complex molecules could not be explained without a proper consideration 
of interstellar dusts \citep{hase92,das10,boog04,gibb04}. Several attempts were made over 
the past few years to study physical and chemical processes on the interstellar grains 
\citep{chak06a,chak06b,cupp07,das08b,das10,das13b,sahu15}.
Several experimental results have been reported in explaining the importance 
of various surface processes on interstellar dusts \citep{iopp08,ober09},
making it imperative to incorporate the grain surface chemistry extensively while predicting their abundances.

The problem of Origin of life is a long standing puzzle and formation of amino acids in the laboratory 
by well known work of \citet{mille53} ushered a new direction of research in this area. More recently,
\citet{chak00a,chak00b} for the first time, suggested that perhaps the process of formation of the 
complex molecules such as Adenine and other constituents of DNA is very generic and such complex pre-biotic 
molecules should be formed during any star forming process. 
In the absence of proper reaction cross sections, \citet{chak00a} used neutral-neutral reaction rates
to compute adenine abundance with successive addition of $\mathrm{HCN}$. This was later improved upon 
with more realistic cross sections \citep{chak00b} and more realistic abundance was obtained. 
Presence of interstellar grains 
were in explicitly added in these works, and were incorporated indirectly by a higher rate coefficients 
in H$_2$ formation. A follow up study by 
\citet{maju12} explicitly considered presence of grains and showed that this significantly 
alters the abundance of adenine. They carried out several prescriptions of 
rate coefficients for successive $\mathrm{HCN}$ addition 
reactions. They also considered radical-radical/radical-molecular reactions proposed by
\citet{gupt11} for the formation of adenine. Their results suggest that radical-radical/
radical-molecular reactions dominates over the neutral-neutral reactions.
Recently \citet{merz14} used the concept of retro-synthetic analysis to produce interstellar 
adenine from observed interstellar molecules such as $\mathrm{C_3NH}$, $\mathrm{HNCNH}$ and its isomer 
$\mathrm{H_2NCN}$. They used MP2/6-311++G(2d,2p) method to calculate various chemical 
parameters involving a six step mechanism. This motivated us to
to perform a comparative study among various available pathways for the formation of adenine 
in interstellar region. This we present below.

The plan of this paper is the following. In Section 2, computational methods are discussed. 
Results are presented in Section 3, and finally in Section 4, we draw our conclusions.

\begin{table*}
\centering
\vbox{
\scriptsize{
\caption{Available and estimated rate co-efficients for the formation of interstellar adenine
in the gas phase via various reaction pathways}
\begin{tabular}{|c|c|c|}
\hline
{\bf Pathways used in}&  {\bf Reactions}&{\bf Rate coefficients} \\
\hline
\hline
&(i)$\mathrm{HCN+HCN \rightarrow CH(NH)CN}$& $8.38 \times 10^{-20} $cm$^3$ sec$^{-1}$\\
{\bf \citet{chak00a,chak00b},}&$(ii)\mathrm{CH(NH)CN+HCN\rightarrow NH_2CH(CN)}$$_2$& $3.43 \times 10^{-12}$ cm$^3$ sec$^{-1}$\\
{\bf \citet{maju12}}&(iii)$\mathrm{NH_2CH(CN)_2+HCN \rightarrow NH_2(CN)C=C(CN)NH_2} $& $3.30 \times 10^{-15} $cm$^3$ sec$^{-1}$\\
&(iv)$\mathrm{NH_2(CN)C=C(CN)NH_2+HCN \rightarrow C_5H_5N_5}$& $3.99 \times 10^{-10} $cm$^3$ sec$^{-1}$\\
\hline
&(v)$\mathrm{HCCN+HCN \rightarrow Molecule \ 1}$& $ 2.13 \times 10^{-17} $cm$^3$ sec$^{-1}$\\
&(vi)$\mathrm{Molecule \ 1 +H \rightarrow Molecule \ 2}$ & $7.96 \times 10^{-9}$ cm$^3$ sec$^{-1}$\\
&(vii)$\mathrm{Molecule \ 2 +NH_2CN \rightarrow Molecule \ 3}$  & $6.24 \times 10^{-12}$ cm$^3$ sec$^{-1}$\\
{\bf \citet{gupt11}}&(viii)$\mathrm{Molecule \ 3 + CN \rightarrow Molecule \ 4}$  & $1.80 \times 10^{-9} $cm$^3$ sec$^{-1}$\\
&(ix)$\mathrm{Molecule \ 4 +H \rightarrow Molecule \ 5}$& $8.71 \times 10^{-9} $cm$^3$ sec$^{-1}$\\
&(x)$\mathrm{Molecule \ 5 + CN \rightarrow C_5H_5N_5+HNC}$& $1.89 \times 10^{-9} $cm$^3$ sec$^{-1}$\\
&(xi)$\mathrm{Molecule \ 5 + CN \rightarrow C_5H_5N_5+HCN}$& $1.91 \times 10^{-9} $cm$^3$ sec$^{-1}$\\
\hline
&(xii)$\mathrm{C_3NH+HNCNH \rightarrow C_4N_3H_3}$& $8.30 \times 10^{-20} $cm$^3$ sec$^{-1}$\\
&(xiii)$\mathrm{C_4N_3H_3+HNCNH \rightarrow C_4H_3N_3 + NH_2CN}$& $6.43 \times 10^{-22} $cm$^3$ sec$^{-1}$\\
{\bf \citet{merz14}}&(xiv)$\mathrm{C_4H_3N_3+HNCNH \rightarrow C_5N_5H_5}$& $1.36 \times 10^{-9} $sec$^{-1}$ \\
&(xv)$\mathrm{C_5N_5H_5+HNCNH \rightarrow N_5C_5H_5 + NH_2CN}$& $1.00 \times 10^{-9} $cm$^3$ sec$^{-1}$\\
&(xvi)$\mathrm{N_5C_5H_5+HNCNH \rightarrow C_5H_5N_5 + NH_2CN}$& $1.89 \times 10^{-13} $sec$^{-1}$\\
\hline
\end{tabular}}}
\end{table*}


\section{Methods and Computational Details}

\subsection{Chemical modeling}

\begin {figure}
\centering
\includegraphics[height=8cm,width=8cm]{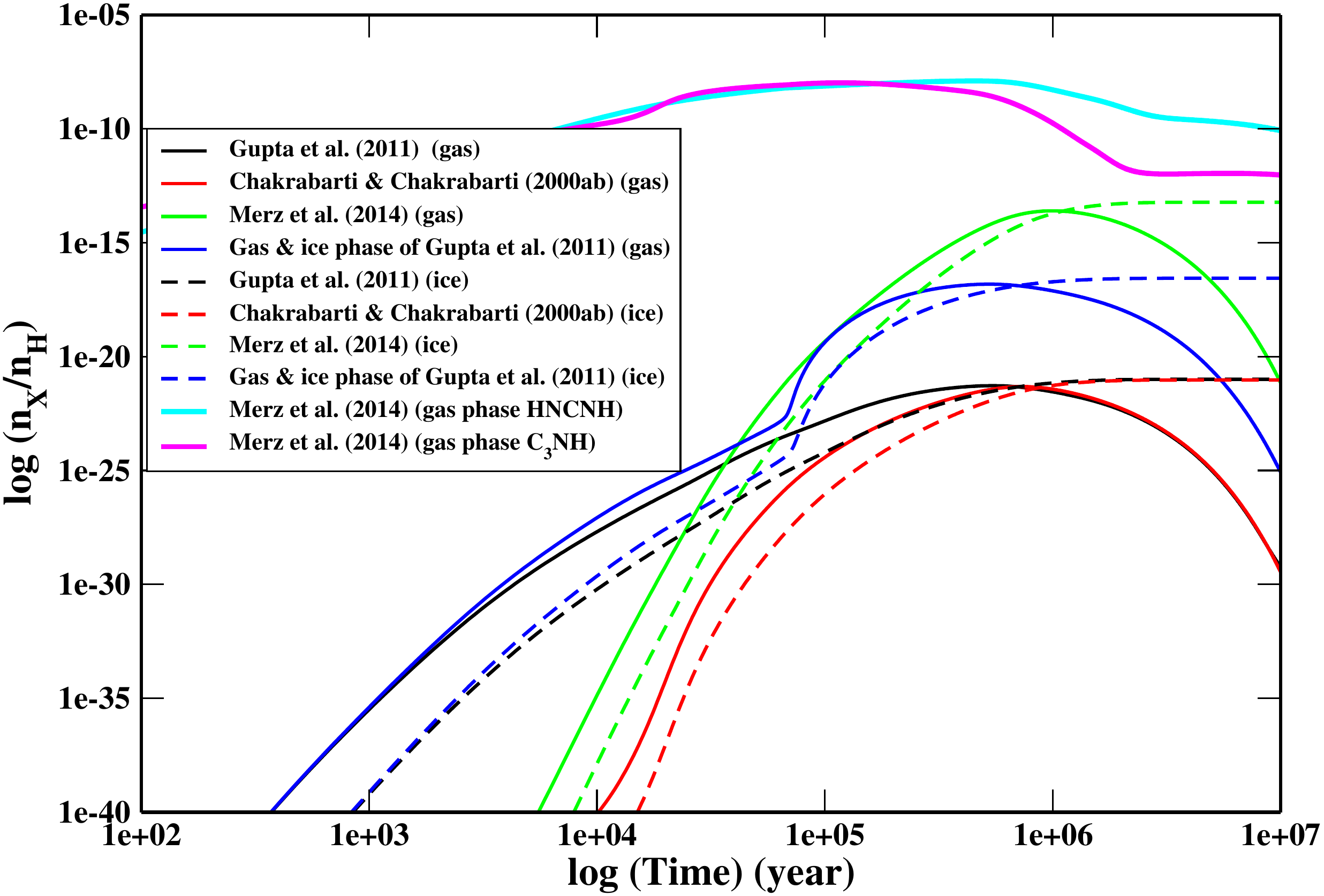}
\caption{\scriptsize Comparison of results of various adenine formation pathways. Pathways proposed
by \citet{merz14} is dominating over all other pathways.
}
\label{fig-1}
\end {figure} 
We develop a large gas-grain chemical network to explore chemical evolution of interstellar adenine 
and its related species. For the gas-phase chemical network, we follow UMIST 
2006 data base. Formation of adenine via various reaction pathways 
used in \citet{chak00a,chak00b}, \citet{gupt11} and \citet{merz14} are included. 
Our gas phase chemical network consists of more than $6000$ reactions between $650$ species.

For the grain surface reaction network, we follow \citet{hase92,cupp07,das10,das11,das13b,das15}. 
Our surface reaction network contains $292$ reactions.  
We consider gas-grain interaction to frame the actual interstellar scenario. Gas phase species are 
allowed to accrete on interstellar grains. Grains are actually acting as a catalyst
for the formation of several complex interstellar species. Formation of the simple and most 
abundant interstellar molecule, $\mathrm{H_2}$ cannot not be explained without the consideration of 
interstellar grain chemistry. Binding energies of surface species mainly dictate chemical composition 
of interstellar grain mantle. We consider the most updated interaction barrier energies as mentioned
in \citet{das13b} and references therein. Depending on the energy barriers, surface species 
would move throughout the grain surface by thermal hopping or tunneling whichever is faster.
At low temperatures, for lighter species such as H atom, tunneling is much faster. 
During these movements, surface species would react with any suitable reactant partners. 
Based on their energy barriers, surface 
species would also thermally evaporate \citep{hase92} and populate the gas phase. 
Cosmic ray induced evaporation \citep{hase93}
and non-thermal desorption \citep{garr07,das15} mechanisms 
are also considered in our model.
So, in brief, gas and grains are interacting with each other to exchange 
their chemical components by various means. Since all the processes are random, Monte Carlo method would 
be appropriate to use while dealing with this randomness. However, it requires huge computational time 
to compute with our vast gas phase and grain phase chemical networks simultaneously. 
Thus, we use traditional Rate equation
method to handle our chemical network on grains. Detailed discussions on our gas-grain 
chemical model are already presented in \citet{das13b,das15,maju14a,maju14b}.

All the reactions, which are considered here for the formation of adenine
are shown in Table 1. \citet{maju12,maju13} considered neutral-neutral 
pathways (reaction number (i)-(iv) of Table 1) of \citet{chak00a,chak00b} and
radical-radical/radical-molecular pathways (reaction numbers (v)-(xi)) of \citet{gupt11}
and concluded that the production of 
adenine is dominated by radical-radical/radical-molecular reaction pathways.
They identified $\mathrm{HCCN}$ and $\mathrm{NH_2CN}$ as two main
precursor molecules, which are responsible for the production of adenine. 
$\mathrm{HCCN}$ is highly abundant in the interstellar space \citep{jiur06,guel91}. 
Formation of $\mathrm{HCCN}$ on the interstellar grains were also 
considered \citep{hase92}. \citet{mcgo96} conducted a deep search 
for $\mathrm{HCCN}$ towards TMC-1 and several GMC's via its $N(J)=1(2) \rightarrow 0(1)$ transition. 
They came up with an upper limit of fractional abundance with respect to molecular 
hydrogen to be $\sim 2 \times 10^{-10}$.
Existence of $\mathrm{NH_2CN}$ in the interstellar cloud was first reported by \citet{turn75}. 
Subsequently, it was observed in both diffuse and dense clouds by \citet{lisz01}. 
\citet{wood07} predicted a steady state fractional abundance  of $2.02 \times 10^{-10}$ 
for $\mathrm{NH_2CN}$ with respect to $\mathrm{H_2}$.

\citet{merz14} proposed a new pathway (reaction nos. x(ii)-(xvi) of Table 1) for the 
production of adenine. They used retro synthetic analysis by using two new species ($\mathrm{C_3NH}$ 
and $\mathrm{HNCNH}$). Carbodimide ($\mathrm{HNCNH}$), is an isomer of cyanamide ($\mathrm{NH_2CN}$).
\citet{mcgu13} obtained the abundance of this molecule from ice mantle experiments. 
They proposed that tautomerization of $\mathrm{NH_2CN}$ on dust grain ice mantles is the dominant
formation pathway for $\mathrm{HNCNH}$. They found its abundance would be $\sim 10\%$ of 
$\mathrm{NH_2CN}$ (in Sgr B2(N) column density of $\mathrm{NH_2CN}$ is $\sim 2 \times 10^{13}$ cm$^{-2}$). 
Since, this was below the detection limit of any current astronomical facility, they proposed to
observe $\mathrm{HNCNH}$ by those transitions which are amplified by masing. In our chemical model, 
we consider that $10\%$ of $\mathrm{NH_2CN}$ could be converted into $\mathrm{HNCNH}$.

We use semi-empirical relationship developed by \citet{bate83} for the computation of 
rate coefficients of the exothermic and barrier less reaction
pathways (reaction nos. xii-xiv) of \citet{merz14}.
\begin{equation}
K = 1 \times 10^{-21} A_r (6E_0 + N -2)^{(3N-7)}/ (3N-7)!  cm^3 s^{-1}
\end{equation}
where, $E_0$ is the association energy in $eV$, $A_r$ is the transition probability 
(in $s^{-1}$) of the stabilizing transition (the numerical value of which may be taken to be $100$ 
unless better information is available) and $N$ is the number of nuclei in reactants. For the reaction 
numbers xii, xiii \& xiv, we use association energies of $-5.2, \ -0.6, \, -72.8 $ kcal/mol 
respectively.

If the calculated rate coefficients from Eqn. 1 exceeds the limit set by the following equation
(Eqn. 2), then this limiting value was adopted:
\begin{equation}
K = 7.41 \times 10^{-10} \alpha^{1/2}(10/\mu)^{1/2} cm^3 s^{-1}
\end{equation}
where, $\alpha$ is the polarizability in {A$^{\circ}$}$^3$, $\mu$ is the reduced mass of the reactants 
in $^{12}C$ amu scale as suggested by \citet{bate83}. 

Rate coefficients of reactions having activation barriers (reaction nos. xv-xvi) 
are calculated by using conventional transition
state theory. According to this theory, the rate coefficient has the following form:
\begin{equation}
k(T) = (K_B T /h C_0 )exp(-\Delta G/RT ) \ s^{-1} ,
\end{equation}
where, $K_B$ is the Boltzmann constant, $h$ is the Plank's constant, $T$ is the temperature,
$C_0$ is the concentration (set to 1, by following \citet{jalb08}), $R$ is the ideal
gas constant, and $\Delta G$ is the free energy of activation. From the quantum chemical calculations
by \citet{merz14},
we are having $\Delta G =+0.4$ kcal/mol and $+1.1$ kcal/mol respectively for reaction numbers xv and xvi.

\begin{table*}
{\centering
\scriptsize
\caption{Initial elemental abundances}
\begin{tabular}{|c|c|}
\hline
{\bf Species}&{\bf Abundance}\\
\hline
\hline
H$_2$ &    $5.00 \times 10^{-01}$\\
He    &    $1.00 \times 10^{-01}$\\
N     &    $2.14 \times 10^{-05}$\\
O     &    $1.76 \times 10^{-04}$\\
H$_3$$^+$&    $1.00 \times 10^{-11}$\\
C$^+$ &    $7.30 \times 10^{-05}$\\
S$^+$ &    $8.00 \times 10^{-08}$\\
Si$^+$&    $8.00 \times 10^{-09}$\\
Fe$^+$&    $3.00 \times 10^{-09}$\\
Na$^+$&    $2.00 \times 10^{-09}$\\
Mg$^+$&    $7.00 \times 10^{-09}$\\
P$^+$ &    $3.00 \times 10^{-09}$\\
Cl$^+$&    $4.00 \times 10^{-09}$\\
e$^-$ &    $7.31 \times 10^{-05}$\\
HD &   $ 1.6 \times 10^{-05}$\\
\hline
\end{tabular}}
\end{table*}

\subsection{Spectroscopic Modeling}

An educated estimate of spectral properties is essential before observing any unidentified species. 
It is reported in earlier astrophysical literature that DFT and TDDFT \citep{rung84} 
computation methods would be applied to various astrophysical problems \citep{pule10,piev14}. 
\citet{das15} discussed the necessity of quantum chemical 
calculations prior to any spectroscopical survey. 
Since our chemical model would predict the abundances of adenine and its related species, 
we believe it would be useful to present various spectral aspects of these species. For finding the
various spectral aspects, we perform quantum chemical calculations by using Gaussian 09W program.
By following \citet{maju14b}, for the vibrational frequencies, we use 
 B3LYP/6-311++G(d,p) level of theory. In case of
grain phase species, Polarizable Continuum Model (PCM) is used with the integral equation 
formalism variant (IEFPCM) as the default SCRF method. 
For the electronic absorption spectrum, we
use time dependent density functional theory (TDDFT study). Since most of the gas phase complex 
molecules were observed by their rotational transitions in the 
mm or sub-mm regime, we carry out quantum chemical calculations to find out 
the rotational transitions of our desired species. 
Computation of the anharmonic frequencies require the use of analytic second 
derivative of energies at the displaced geometries. But the CCSD method 
in Gaussian 09W program only implements energies, so analytic second derivative 
of energies are not available at this level of theory. So there are no options in 
Gaussian 09W program to compute the rotational and distortional constants at the 
CCSD level of theory. Here, we use B3LYP/aug-cc-pvTZ level of theory, which are also 
proven to be very accurate \citep{carl13,das15}.
This level of theory would be appropriate because there are some earlier studies 
(such as, \citet{brun06}) for finding rotational and distortional constants of Uracil 
(one nucleo base of RNA beside adenine, cytosine and guanine).
Corrections for the interaction between the rotational motion and
vibrational motion along with the corrections for vibrational averaging and
an-harmonic corrections to the vibrational motion are also considered in our calculations.
Obtained rotational and distortional constants are then used in
SPCAT \citep{pick91} program to predict various rotational transitions.
Output of the SPCAT program is then directly used in ASCP Prgram \citep{kisi98,kisi00}
to find out the rotational stick diagram of the desired species.

\section{Results and Discussion}

\subsection{Results of Chemical modeling}

\begin {figure}
\centering
\includegraphics[height=8cm,width=8cm]{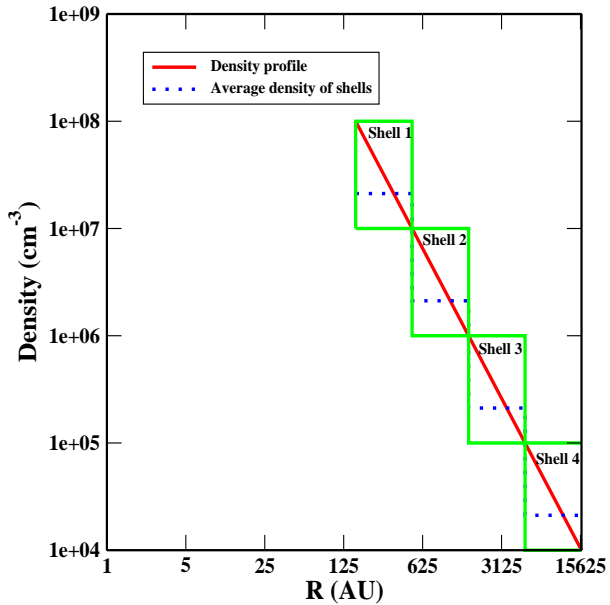}
\caption{\scriptsize Radial distribution of density profile.}
\label{fig-2}
\end {figure} 
\citet{chak00a,chak00b,maju12} considered successive 
$\mathrm{HCN}$ addition for the formation of adenine. \citet{gupt11} proposed adenine 
formation pathways starting with $\mathrm{HCCN}$ and $\mathrm{HCN}$. \citet{gupt11} 
also pointed out that their reaction pathways would also be feasible in the grain phase. 
A completely new pathway has been proposed recently by \citet{merz14}, where
adenine could be produced without any $\mathrm{HCN}$ addition. 
So, we wish to compare four different pathways where the first one
is the gas phase pathway of \citet{gupt11}, the second one
is the pathway considered by \citet{chak00a,chak00b} 
and updated by \citet{maju12}, the third one is the gas phase pathway proposed by \citet{merz14}. As a fourth method,
we consider both gas and grain phase pathways of \citet{gupt11}. 
Since, no mention was made of the grain phase reactions in \citet{merz14} 
and \citet{chak00a,chak00b}, we do not consider these pathways in the grain phase. 
In general, we consider that the gas and the grain interact with each other to exchange their 
chemical components. So, grain phase could be populated by the acrretion from the gas phase and 
gas phase could be populated by the grain phase species through evaporation mechanisms 
such as, thermal desorption, non-thermal desorption 
and cosmic-ray induced desorption.
Initial elemental abundances with respect to total hydrogen nuclei (Table 2) are considered 
to be similar to what is in 
\citet{das13b} and \citet{maju14b}. These are the typical
low metallic abundances which are often adopted for TMC-1 cloud. 
In Fig. 1, a comparative study between these four pathways are shown 
for the chemical evolution of adenine. 
To draw this, we take number density of hydrogen nuclei existing in all possible forms 
($n_H$) to be $= 10^4$ cm$^{-3}$, 
$A_V=10$ and $T=10$ K. Various evaporation mechanisms are considered here, among them, around 
low temperatures, non-thermal desorption mechanism \citep{garr07,das15} is 
the most efficient means to transfer these complex molecules to the gas phase. 
A moderate value of the desorption parameter ($\sim 0.05$) is considered. 
Significant difference in the abundance of adenine could be observed between the 
two considerations of \citet{gupt11}. In one case, only gas phase pathways of 
\citet{gupt11} is considered and in another case, the same pathways
are considered for both the phases (gas and grain). From Fig. 1, we find that the peak 
abundances of adenine for these 
two cases appear to be $5.3 \times 10^{-22}$ and $1.5 \times 10^{-17}$ respectively. 
Since grain phase production is efficient, production is 
significantly enhanced in the case when grain phase pathways are also considered. 
While using pathways of \citet{chak00a,chak00b}, peak abundance 
of adenine turns out to be $4.8 \times 10^{-22}$ 
and using pathways proposed by \citet{merz14}, the peak abundance becomes $2.5 \times 10^{-14}$.
Using neutral-neutral rate coefficients in an isothermal cloud, 
\citet{chak00a} found the abundance of $\sim 5 \times 10^{-13}$, one magnitude higher than this. 
It is also clear from Fig. 1 that the adenine formation pathways of \citet{merz14} dominates over 
all other pathways. In Fig. 1, we show chemical evolutions of $\mathrm{HNCNH}$ and $\mathrm{C_3NH}$ 
(which are required for the formation of adenine in the pathways proposed by \citet{merz14}). 
These molecules could be used as the precursors for estimating the abundances of adenine 
in interstellar region.

\subsubsection{Mixing model}

Chemical complexity of any interstellar region is highly dependent on surrounding physical 
condition. Formation of bio-molecules would also be influenced by surrounding physical process.
To describe a realistic situation, we prepare a special model named ``Mixing model".
We consider that density profile follows $\rho \sim r^{-2}$ 
distribution as described by \citet{shu77}. At the outer boundary of the cloud 
the density is $10^4$ cm$^{-3}$, inner boundary location is chosen in such way that the
density would become $10^8$ cm$^{-3}$ (Fig. 2). For concreteness, we choose the 
outer boundary to be at $15962$ AU and the inner boundary becomes at $159.62$ AU. 
To reduce computational time we further subdivide this cloud into four shells, 
innermost shell is extended from $159.62$ AU to $504.76$ AU which has an average density of 
$2.12 \times 10^7$ cm$^{-3}$. The shell no. $2$ has an average density of 
$2.12 \times 10^6$ cm$^{-3}$ and extends from $504.76$ AU to $1596.2$ AU.
The shell no. $3$ has an average density of $2.12 \times 10^5$ cm$^{-3}$ and extends 
from $1596.2$ AU to $5047.63$ AU. The fourth shell 
extends from $5047.63$ AU to $15962$ AU and has an average density of $2.12 \times 10^4$ cm$^{-3}$. 
Since the cloud is dense, we assume visual extinction ($A_V$) of $10$ and 
temperature $10$ K throughout the cloud.

\begin {figure}
\centering
\includegraphics[height=7cm,width=7cm]{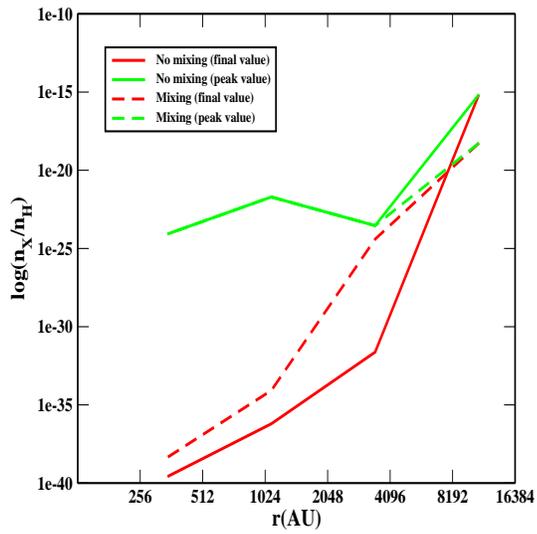}
\caption{\scriptsize Radial distribution of adenine when mixing and no mixing models are considered. 
Due to heavy mixing
the final and peak abundances (dashed lines) become similar.}
\label{fig-3}
\end {figure} 

We assume that a systematic mixing is going on throughout the cloud. Matter of shell 4 would contribute 
($p_4 \%$ of shell 4 matter) to shell 3, material of shell 3 would contribute ($p_3 \%$ of shell 3 matter) 
to shell 2, material of shell 2 would contribute ($p_2 \%$ of shell 2 matter) to shell 1 and due to 
re-entry of the outflow at the outer edge, some matter would re-enter 
($p_1 \%$ of shell 1 matter) into shell 4. Each transport should have different 
time scales. Inward material transfer (from shell 4$\rightarrow$ shell 3$\rightarrow$
shell 2 $\rightarrow$shell 1) is assumed to take place with the sound speed of ($\sqrt{\gamma KT/m_p}\sim 
2.873 \times 10^4$ cm/s for $T=10$ K and $\gamma=5/3$). Since shell 4 is $10914 \ AU$ thick, sound wave will 
take $5.68 \times 10^{12} \ s$ ($t_4$) to cross it. We assume that after every $t_4$ second, some percentage 
matter would move to shell 3. Similarly, we calculate the time scale for transferring some fraction of
matter from shell 3 to shell 2 ($t_3=1.80 \times 10^{12} \ s$) and shell 2 to shell 1 
($t_2=5.68 \times 10^{11} \ s$). Due to outflow, some fraction of matter of shell 1 would 
contribute to shell 4. Time scale for the outflow ($t_1$) is chosen by considering free fall time scale
($\sqrt{3\pi/32G\rho}$). Free fall time scale is inversely proportional to the square root 
of density. Here, we choose a typical dense cloud number density ($10^4 \ cm^{-3}$) 
for the calculation of free fall time. So in our case we assume $t_1=1.62 \times 10^{13} \ s$. For
simplicity, we assume that $p_4=p_3=p_2=p_1=20\%$.

In Fig. 3, depth dependence of peak and final adenine abundances are shown
for the non-mixing and mixing cases with a density profile of Fig. 2. 
Abundances are represented with respect to the total hydrogen nuclei in all forms.
For each shell, We carry out our simulation up to $2 \times 10^6$ year. By the 
peak abundance, we mean the peak value obtained during the simulation regime and 
by the final abundance, we mean the value obtained at the end of simulation time scale.
Since from Fig. 1, it is clear that the pathways of \citet{merz14} is dominating 
over all other existing pathways for the production of adenine in interstellar region,
here, for Fig. 3, we considered only the pathways of \citet{merz14}.
Clearly, due to the mixing of matter 
among various shells, final abundance of adenine is severely affected in the mixing model. Peak abundances
of adenine is decreasing inside the cloud. Reason behind is that as we penetrate 
inside the cloud, density increases and this results in shorter depletion (to the next shell) time scale. 
Due to heavy depletion, all related species depleted from the gas phase. As a result,
the peak value of adenine is reduced. However, the final value goes up and indeed, the final and the peak 
values become similar due to significant mixing. 

\subsection {Results of Spectroscopic modeling}

\subsubsection{Vibrational transitions}
$\mathrm{HNCNH}$ is one of the isomers of $\mathrm{NH_2CN}$ which is a planar pentatonic molecule. 
It has nine fundamental vibrations \citep{rubl56}. Similar to $\mathrm{NH_2CN}$, $\mathrm{HNCNH}$ 
also has nine fundamental vibration lines as shown in Table 3. In Fig. 4a-c, infrared absorption spectra of 
$\mathrm{HNCNH}$ for the gas phase and grain phase are shown. We note that the gas phase
$\mathrm{HNCNH}$ has its strongest feature at $2225.48$  cm$^{-1}$ and the second intense peak 
arises at $899.71$  cm$^{-1}$. In the grain phase $2225.48$  cm$^{-1}$ peak is shifted to 
$2179.37$ cm$^{-1}$ and peak at $899.71$  cm$^{-1}$ is also shifted to $925.14$  cm$^{-1}$ with 
much lower intensity. Beside these intense peaks, there are several peaks which are pronounced in 
our grain phase model (such as peaks at $902.61$  cm$^{-1}$ and $3564.13$  cm$^{-1}$). For the sake of 
better understanding, we placed all the infrared peak positions with their absorbance in Table 3. 
Moreover, a comparison with the other theoretical/experimental vibrational transitions 
for $\mathrm{HNCNH}$ (Birk \& Winnewisser, 1986; King \& Strope, 1971) are also
shown in Table 3. Some of our peak positions are very close to the results reported in the literature.

\begin {figure}
\centering
\includegraphics[height=6cm,width=6cm]{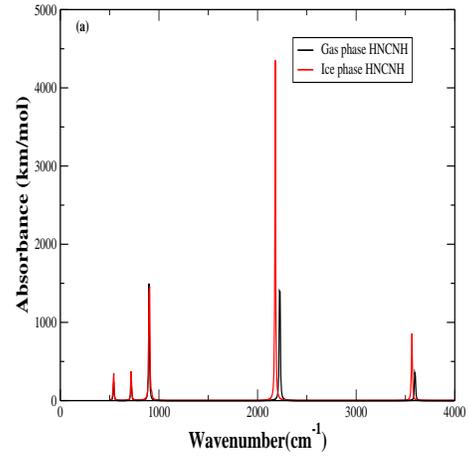}
\vskip 1cm
\includegraphics[height=6cm,width=6cm]{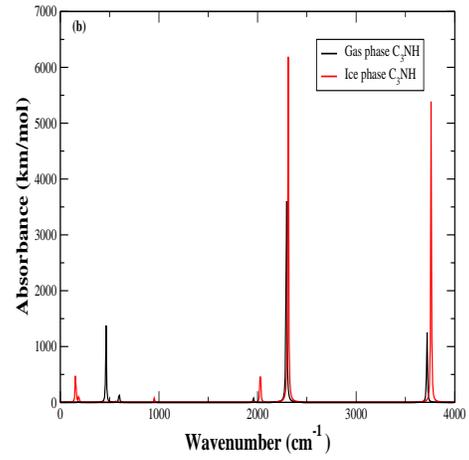}
\vskip 1cm
\includegraphics[height=6cm,width=6cm]{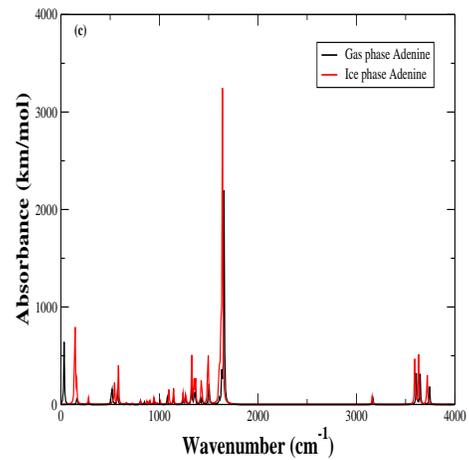} 
\caption{\scriptsize Infrared spectra of $\mathrm{HNCNH}$, $\mathrm{C_3NH}$ and adenine in gas and grain (water ice) phases.}
\label{fig-2}
\end {figure} 

In Table 3, we also note down the peak position of the infrared spectrum of $\mathrm{C_3NH}$ in the 
gas as well as in the grain phase. We find that the most intense mode in the gas phase appears at 
$2293.10$  cm$^{-1}$. This peak is shifted in the left side in the grain phase and appears at 
$2309.68$  cm$^{-1}$. Among the other peaks in the gas phase, $3718.05$  cm$^{-1}$ and 
$463.33$  cm$^{-1}$ has the major contributions.
One strong peak at $3760.19$  cm$^{-1}$ appears in the grain phase.  
Fig. 4c shows infrared spectra of adenine in gas 
and grain phases. Gas phase infrared spectrum of adenine contains strong peaks at $1634.40$  cm$^{-1}$, 
$1656.77$  cm$^{-1}$, $31.23$  cm$^{-1}$ and all these peaks are shifted to grain phase and appears at 
$1638.84$  cm$^{-1}$, $1621.56$  cm$^{-1}$, $141.19$  cm$^{-1}$. Not only that, few new strong peaks 
appears at $3594.23$  cm$^{-1}$, $3634$  cm$^{-1}$, $3718.16$  cm$^{-1}$ for grain phase adenine. 
These features are clear in the Fig. 4c. 

\subsubsection{Electronic transitions}
We continue our computation to obtain the spectral properties of $\mathrm{C_3NH}$ and $\mathrm{HNCNH}$ in the
electronic absorption mode. Electronic absorption spectra of $\mathrm{HNCNH}$, $\mathrm{C_3NH}$ and
adenine in gas as well in the grain phase are shown in Fig. 5a-c. Corresponding electronic transitions, 
absorbance and oscillator strengths are also summarized in Table 4. An electronic absorption spectrum of 
$\mathrm{HNCNH}$ molecule in the gas phase is characterized by five intense peaks. These five 
transitions occurred at $219.55$, $184.83$, $156.97$, $141.48$, $102$ $nm$ with major contributions 
from  1-A1 $\rightarrow$1-B2, 1-A1$\rightarrow$1-B1, 1-A1$\rightarrow$1-B2, 1-A1$\rightarrow$1-B2 and 
1-A1$\rightarrow$ 1-A2 electronic transitions. All the peaks are shifted in grain phase with the
corresponding change in electronic transitions, absorbance and oscillator strength. 
For $\mathrm{C_3NH}$ molecule, electronic absorption spectra in gas as well in the grain 
phases are shown in Fig. 5b. This electronic spectrum in gas phase contains peaks at 
$182.6$, $126.7$, $104.04$ nm due to $1-A' \rightarrow1-A'$, $1-A'\rightarrow1-A"$,
$1-A' \rightarrow 1-A'$. As in case of $\mathrm{HNCNH}$, here also corresponding gas phase peaks 
shift in the grain phase. However, most interestingly, the gas phase as well as grain 
phase adenine contains only one strong peak in the electronic mode shown in the Fig. 5c. Gas phase 
strong peak appears at $256.12$ nm with oscillator strength $0.0655$ for $1-A\rightarrow 1-A$ 
electronic transition. This peak is shifted to grain phase and appears at $252.11$ nm with oscillator 
strength $0.1267$ for the same electronic transition.

\subsubsection{Rotational transitions}
In Table 5, we summarize our calculated rotational and distortional constants of the precursors
$\mathrm{HNCNH}$ and $\mathrm{C_3NH}$ of adenine  and adenine itself,
in gas phase. Calculated constants are corrected for each vibrational 
state as well as vibrationally averaged structures. Here we use B3LYP/cc-pVTZ level to 
perform these calculations in gas phases. In Table 5, calculated distortional constants correspond to 
I$^t$ representation with `A' reduction. Figure 6 shows the stick 
diagram of two new precursors of adenine. 

\begin {figure}
\centering 
\includegraphics[height=6cm,width=6cm]{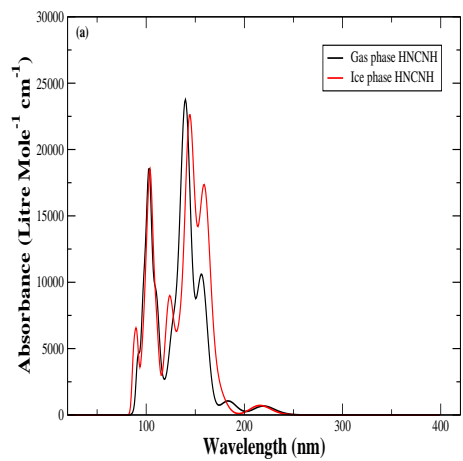}
\vskip 1cm
\includegraphics[height=6cm,width=6cm]{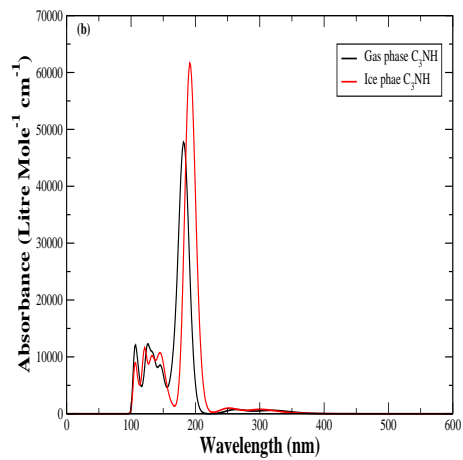}
\vskip 1cm
\includegraphics[height=6cm,width=6cm]{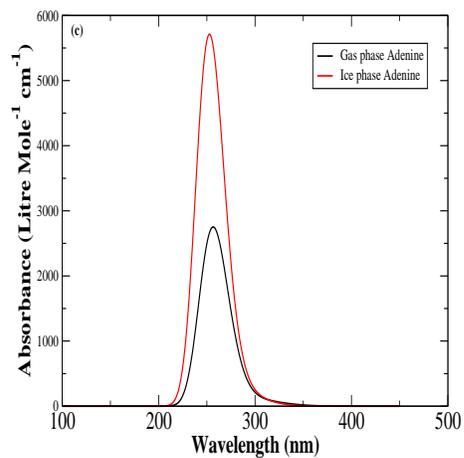}
\caption{\scriptsize Electronic absorption spectra of $\mathrm{HNCNH}$, $\mathrm{C_3NH}$ and
adenine in gas and grain (water ice) phases.}
\label{fig-4}
\end {figure}

\begin{table*}
\scriptsize{
\centering
\vbox{
\caption{Vibrational frequencies of precursors HNCNH and C$_3$NH 
of adenine in gas phase and H$_2$O ice containing grains at B3LYP/6-311G++(d,p) level}
\begin{tabular}{|c|c|c|c|c|c|}
\hline
{\bf Species}&{\bf Peak positions }&{\bf Absorbance}&{\bf Peak positions }&{\bf Absorbance}&{\bf Experiment or other}\\
{}&{\bf (Gas phase)}&{}&{\bf (water ice)}&{}&{\bf theoretical value}\\
 &{\bf (Wavenumber in $cm^{-1}$)}&& {\bf (Wavenumber in $cm^{-1}$)}&& {\bf (Wavenumber in $cm^{-1}$)}\\
\hline
&540.31& 79.926 & 539.48 & 141.348&$537^k$\\
& 540.65 & 0.195 & 539.60 & 0.142 &\\
& 718.17 & 107.962 & 719.18 & 140.079 &\\
&897.11 & 9.975 & 902.61 & 712.664 &890 $\pm$ 10$^b$, 886$^k$ \\
{\bf HNCNH}& 899.71  & 438.296  &925.14  &  18.842  &\\
& 1287.51 & 0.0593 & 1284.96 & 0.3371 &1285 $\pm$ 20 $^k$, 1275$^b$\\
& 2225.48 &  713.448 & 2179.37 & 1290.062& 2104.7$^b$, 2097$^k$\\
& 3596.90 & 143.912& 3564.13& 269.387 &	\\
& 3599.41 & 26.106 & 3569.93 & 52.647 &\\
\hline
& 180.31 & 3.324 &154.87  & 207.6288  &\\
& 190.52 & 1.491 &  187.05  & 38.291    & \\
& 463.33 & 409.108 & 188.75   & 2.815  & \\
& 592.72 & 2.186 & 563.39   & 8.437    &\\
{\bf C$_3$NH}  & 596.46 & 62.223 &  572.20  & 2.130   & -\\
& 965.33 & 0.683 & 953.06   & 23.805  &\\
& 1957.85 & 27.241 & 2028.00   & 259.828   &\\
& 2293.10 & 1589.606 & 2309.68 & 2393.2106   & \\
& 3718.05  &  448.645 & 3760.19 & 1561.688  &  \\
\hline
\multicolumn{6}{|c|}{$^b$ Birk and Winnewisser, 1986}\\
\multicolumn{6}{|c|}{$^k$ King and Strope, 1971}\\
\hline
\end{tabular}}}
\end{table*}

\begin{table*}
\centering
\addtolength{\tabcolsep}{-4pt}
\scriptsize{
\vbox{
\caption{Electronic transitions of precursors HNCNH and C$_3$NH
of adenine at B3LYP/6-311++g(d,p) 
level theory in gas phase and H$_2$O ice containing grain phase}
\begin{tabular}{|c|c|c|c|c|c|c|c|c|}
\hline
{\bf Species}&{\bf Wavelength}&{\bf Absorbance}&{\bf Oscillator strength }&{\bf Transitions}

&{\bf Wave length}&{\bf Absorbance}&{\bf Oscillator strength}&{\bf Transitions}\\

{}&{(gas phase)}&{}&{}&{}&{(H$_2$O ice)}&{}&{}&{}\\
&(in nm)&&&&(in nm)&&&\\
\hline
& 182.6 & 47462.573 & 0.9608& 1-A'$\rightarrow$1-A'& 191.57 & 61654.567 & 1.5193 & 1-A$\rightarrow$1-A\\
& 126.7 & 12259.477 & 0.0073 & 1-A'$\rightarrow$1-A" & 143.78 & 10621.323 & 0.0276 & 1-A$\rightarrow$1-A\\
{\bf C$_3$NH}& 104.04 & 9869.950 & 0.0216 & 1-A'$\rightarrow$1-A'& 119.07 & 11042.926 & 0.0593 & 1-A$\rightarrow$1-A\\
& - & - & -& -& 104.3 & 7174.6216 & 0.0034 & 1-A$\rightarrow$1-A\\
\hline
& 219.55 & 672.854 & 0.0166 & 1-A1$\rightarrow$1-B2& 215.94 & 716.869 & 0.0177 & 1-A1$\rightarrow$1-B2\\
& 184.83 & 1043.151 & 0.0217 & 1-A1$\rightarrow$1-B1 & 159.3 & 17358.816 & 0.4126 & 1-A1$\rightarrow$1-B2\\
{\bf HNCNH}& 156.97 & 10575.579 & 0.2225 & 1-A1$\rightarrow$1-B2& 144.73 & 22602.596 & 0 & 1-A1$\rightarrow$1-A2\\
& 141.48 & 22671.816 & 0.4362 & 1-A1$\rightarrow$1-B2 & 123.29 & 8869.585 & 0.1003 & 1-A1$\rightarrow$1-A1\\
& 102 & 18576.3854 & 0 & 1-A1$\rightarrow$1-A2 & 100.94 & 14075.884 & 0.0338 & 1-A1$\rightarrow$1-B2\\
& - & - & -& -& 87.24 & 5237.201 & 0.1028 & 1-A1$\rightarrow$1-B1\\
\hline
\end{tabular}}
}
\end{table*}

\begin{table*}
\scriptsize{
\centering
\vbox{
\caption{Theoretical rotational parameters of adenine and its two precursors HNCNH and C$_3$NH at
B3LYP/cc-pVTZ level of theory}
\begin{tabular}{|c|c|c|c|c|}
\hline
{\bf Species}&{\bf Rotational }&{\bf Values}&
{\bf Distortional }&{\bf Values} \\
&{\bf constants }&{\bf in MHz}& {\bf  constants }&{\bf in MHz} \\
&&&& \\
\hline
\hline
&A& 378941.67576 & $D_J$& 0.30806$\times$10$^{-2}$ \\
{\bf HNCNH in gas phase}& B & 10550.84447 & $D_{JK}$& 0.36739$\times$10$^{0}$ \\
&C& 10334.08855 & $D_{K}$& 0.11423$\times$10$^{3}$    \\
& & & $d_1$ & $-0.89721\times$10$^{-5}$ \\
&&& $d_2$& $-0.12535\times$10$^{-4}$\\
\hline
&A& 1431947.28270   & $D_J$& 0.48833$\times$10$^{-3}$ \\
{\bf C$_3$NH in gas phase}& B & 4694.90368 & $D_{JK}$& -0.19558$\times$10$^{0}$ \\
&C& 4680.33880 & $D_{K}$& 0.41486$\times$10$^{5}$    \\
& & & $d_1$ & $-0.20704\times$10$^{-5}$ \\
&&& $d_2$& $0.29323\times$10$^{-6}$\\
\hline
&A& 2387.05922 & $D_J$& 0.19731$\times$10$^{-4}$ \\
{\bf C$_5$H$_5$N$_5$ in gas phase}& B &1575.31931  & $D_{JK}$& 0.14017$\times$10$^{-3}$ \\
&C& 949.02096 & $D_{K}$& 0.10519$\times$10$^{-4}$    \\
& & & $d_1$ & $-0.10350 \times$10$^{-4}$ \\
&&& $d_2$& $ -0.49530 \times$10$^{-5}$\\
\hline
\end{tabular}}}
\end{table*}

\begin {figure}
\vskip 2cm
\centering
\includegraphics[height=7cm,width=10cm]{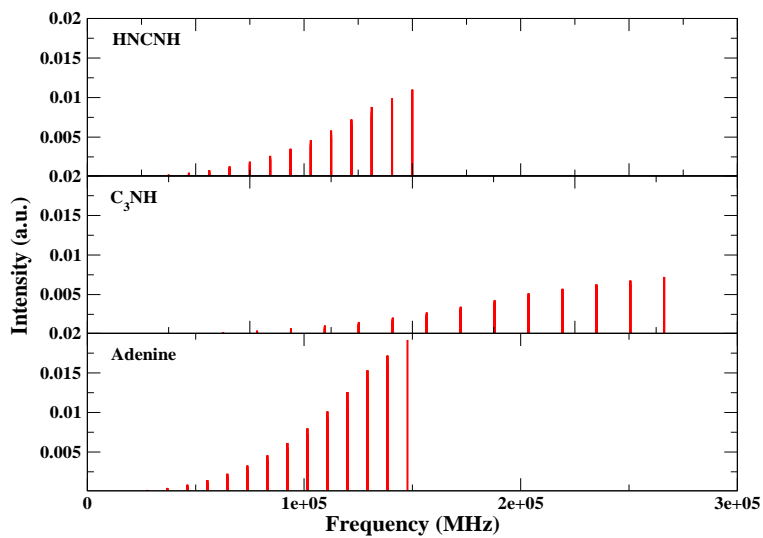}
\caption{\scriptsize Rotational stick diagrams of $\mathrm{HNCNH}$, $\mathrm{C_3NH}$ and adenine.}
\label{fig-7}
\end {figure}

\section {Concluding Remarks}

Interstellar chemistry to form complex bio-molecules began in this century and 
the first quantitative computation of adenine and other pre-biotic molecules 
was presented by \citet{chak00a,chak00b}. A follow up study by
\citet{maju12} showed that formation of adenine mainly follows
radical-radical/radical-molecular reaction pathways as proposed by \citet{gupt11}. 
More recently, \citet{merz14} performed retro synthetic analysis to find out a new mechanism for adenine
formation in the gas-phase. They proposed that it could be formed from three already known interstellar 
molecules, namely, $\mathrm{C_3NH}$ and the isomers, $\mathrm{HNCNH}$ and ${\mathrm H_2NCN}$. 
Uniqueness of this pathway is that it does not involve $\mathrm{HCN}$, water or ammonia 
during the production of adenine by any of the 
six intermediate steps. In this paper, we carried out a comparative 
study among various formation pathways of adenine. We considered all the pathways used 
in \citet{chak00a,chak00b,gupt11,merz14}.
Our chemical model suggests that the adenine formation pathways proposed by \citet{merz14} 
dominates over all other existing pathways. 
We studied the effects of mixing caused by 
both advection and outflows. We find that for adenine, though the peak abundance is severely reduced, the final 
abundance becomes higher as compared to the non-mixing (static evolution) case.
We explored the infrared, electronic, sub-millimeter spectroscopy of adenine along with its two new precursor molecules 
($\mathrm{HNCNH}$ and $\mathrm{C_3NH}$). We show that adenine will have one single strong line either
in the gas or in the grain phase vibrational spectra. A detailed chemical and 
spectroscopical information about adenine
as revealed by our analysis and its precursors would be extremely helpful for
future survey of this species in the interstellar medium. In case the adenine is not directly
detected, we believe that the detection of $\mathrm{HNCNH}$, ${\mathrm H_2NCN}$ and $\mathrm{C_3NH}$ 
would give adequate reason to believe that adenine must also be present.

\section{Acknowledgments}
AD, SKC are grateful to ISRO respond (Grant No. ISRO/RES/2/372/11-12)  and 
DST (Grant No. SB/S2/HEP-021/2013) for financial support. LM thank MOES for funding during this work. 

\clearpage
{\scriptsize

\end{document}